\documentclass[sn-nature,referee, pdflatex]{sn-jnl}% Style for submissions to Nature Portfolio journals
\usepackage{graphicx}%
\usepackage{multirow}%
\usepackage{amsmath,amssymb,amsfonts}%
\usepackage{amsthm}%
\usepackage{mathrsfs}%
\usepackage[title]{appendix}%
\usepackage{xcolor}%
\usepackage{textcomp}%
\usepackage{manyfoot}%
\usepackage{booktabs}%
\usepackage{algorithm}%
\usepackage{algorithmicx}%
\usepackage{algpseudocode}%
\usepackage{listings}%
\usepackage{vmargin}
\usepackage{physics}
\begin{document}

\title[Article Title]{Electron Spin Polarization as a Predictor of Chiroptical Activity in Helical Molecules}

\author*[1]{\fnm{Solmar} \sur{Varela}}\email{solmar.varela@tu-dresden.de}

\author[1]{\fnm{Rafael} \sur{Gutierrez}}
%\email{iiauthor@gmail.com}

\author[1,2]{\fnm{Gianaurelio} \sur{Cuniberti}}
%\email{iiiauthor@gmail.com}

%\author[3]{\fnm{Jesus M}. \sur{Ugalde}}
%\affiliation{Kimika Fakultatea, Euskal Herriko Unibertsitatea (UPV/EHU), P.K. 1072, 20018 Donostia, Spain}

\author[3,1]{\fnm{Ernesto} \sur{Medina}}
%\email{iiiauthor@gmail.com}

\author*[4]{\fnm{Vladimiro} \sur{Mujica}}\email{vmujica@asu.edu}

\affil*[1]{\orgdiv{Institute for Materials Science and Max Bergmann Center of Biomaterials}, \orgname{TU Dresden}, \postcode{01062} \city{Dresden}, \country{Germany}}

\affil[2]{\orgdiv{Dresden Center for Computational Materials Science (DCMS)}, \orgname{TU Dresden}, \postcode{01062} \city{Dresden}, \country{Germany}}

%\affil[3]{\orgdiv{Kimika Fakultatea}, \orgname{Euskal Herriko Unibertsitatea (UPV/EHU), P.K. 1072}, \postcode{20018} \city{Donostia}, \country{Spain}}

\affil[3]{\orgdiv{Departamento de F\'isica, Colegio de Ciencias e Ingenier\'ia}, \orgname{Universidad San Francisco de Quito},  \postcode{170901} \city{Quito}, \country{Ecuador}}

\affil*[4]{\orgdiv{School of Molecular Sciences}, \orgname{Arizona State University}, Tempe, AZ \postcode{85287} \country{USA}}


\abstract{
Chiral structures, breaking spatial inversion symmetry, exhibit non-zero chiroptical activity (COA) due to the interaction between their electric and magnetic responses under external electromagnetic fields, an effect that is otherwise absent in achiral systems. Non-magnetic chiral structures also exhibit Chiral-Induced Spin Selectivity (CISS), where spin-polarization (SP) emerges without external magnetic influence. We have obtained a COA-SP connection for a model system of an electron constrained to a helix including spin-orbit coupling (SOC), and in the presence of an external electromagnetic field. Despite its simplicity, this model captures the relevant physics required to address the problem. In particular, our results reveal that the norm of the SP vector can be used as a predictor of COA. In addition to SOC and the breaking of space inversion, a non-vanishing SP requires the breaking of time-reversal symmetry (TRS), as demanded by Onsager's reciprocity. Beyond the relationship between SP and COA, we obtain the novel result that TRS breaking is also necessary to yield a non-vanishing contribution of the SOC to the COA.}

\keywords{CISS, chiroptical activity, spin-orbit coupling, perturbation theory}



\maketitle

%\section{Introduction}\label{sec1}
Spin polarization (SP) in molecules and solids is usually understood as a response to an external magnetic field. The discovery that for chiral materials, molecules, solids, and interfaces, this magnetic response can be obtained even in the absence of external magnetic  fields in processes involving electron transfer, electron transport, and bond polarization through chiral centers, has had profound consequences in fundamental physics, chemistry, and biology,  as well as in important applications in spintronics, NMR, and Quantum Information Sciences (QIS)\cite{ACSNano2022,Shi2004,Sun2014,Dor2013,Dor2017,Shinto2014,Santos2018,Varade2018,Bustami2020,Bustami2022,Chiesa2023}. The theoretical description of this phenomenon, known as the Chiral-Induced Spin Selectivity (CISS) effect, requires first, the inclusion of spin-orbit coupling (SOC), and second,  in addition to space-inversion symmetry breaking associated with chirality, the breaking of time-reversal  symmetry (TRS) \cite{Wees2019,Wees2020,VarelaScipost2023,Dednam2023}. These basic symmetry requirements do not exclude the fact, recently established in a number of theoretical studies, that a true comprehension of the physics of the CISS effect demands the inclusion of electron-phonon, spin-phonon, and electron-electron interactions in addition to non-adiabatic effects\cite{Volosniev2021,Subotnik2021,Peralta2020,Fransson2023,Peralta2023}. The CISS effect has been extensively investigated in experiments involving photoemission\cite{Ray1999,Vager2010,Gohler2011,Weiss2019}, electron transport\cite{Kettner2015,Xie2011,Nogues2011,Kiran2016}, electron transfer\cite{Malajovich2000,Min2003}, and more recently, in electrochemistry and photoluminescense \cite{Wei2006,Debabrata2013,Mondal2015,Beratan2017,GhoshWaldeck2019,Zwang2018,Torres2020}.

The interpretation of the  CISS effect as a  magnetic response linked to electron SP in chiral systems strongly suggests that there must exist a connection with the optical activity response observed in these systems when exposed to circularly polarized light \cite{Beratan2017,GranadaCISSOptical,PaltierOpticalCISS2023,Intercalators2022,MujicaFieldMediated2020,FranssonTOptical2022}.This connection arises from the interplay between the electric and magnetic induced dipole moments in chiral systems. For optically active molecules, the optical response is normally measured either by the rotational power, which expresses the angle of rotation of the polarization plane of the light, or through the Chiroptical Activity (COA), related to the difference in the intensity of absorption of polarized photons. On the other hand, SP is defined as a three-dimensional vector containing the expectation value of the three Pauli spin matrices, which in the CISS mechanism is associated with electron transport, electron transfer, or bond polarization in chiral systems. Both SP and COA can be interpreted as responses of a chiral system to perturbations, but the underlying physical mechanisms for these two phenomena are fundamentally different because COA requires photon absorption, whereas CISS-SP does not necessarily involve photoexcitation. The latter might explain why the explicit connection between them has so far remained elusive \cite{Xie2022,Kubo2021,Naskar2022}. 

We have used a model Hamiltonian of an electron confined to a helical box in the presence of SOC induced by the confining field of the inversion asymmetric helix \cite{Arraga2015}. In the absence of SOC, the model was initially solved exactly by Tinoco et al. \cite{Tinoco1964}, and  we extend it further  by  including an external electromagnetic field.  Using a perturbative approach to solve the Schr\"odinger equation, we approximately obtain the two components of the $1/2-$spinors for this problem.  We then proceed to calculate both the COA and the three components of the CISS-induced magnetic response, where the $z$-component corresponds to the average value of $\sigma_{z}$ Pauli matrix, assuming electron motion along the $z$-axis. Despite the apparent simplicity of our model, it is the first explicit analytical solution to a long-standing problem in the field of CISS-induced molecular SP and how it can be connected to COA. Our model is also related to the description of the phenomenon of field-mediated chirality transfer ref.\cite{MujicaFieldMediated2020}. The conclusion is that the interaction with circularly polarized radiation carries information both about SP and COA, an important result that we are only beginning to understand and that might have important consequences in the description of SP using chiral photons in molecules with emerging topological features in the electronic structure as, for example, spin textures\cite{Binghai2023}.  

Our model also affords an important digression into the inclusion of controlled schemes to take into account TRS breaking. This is a fundamental aspect of any formalism used to describe SP because it cannot occur in a system where TRS is preserved, as implied by Onsager relations and the onset of Kramers' degeneracy \cite{ButtikerOnsager2012,Bardarson2008}. We have used a simple realization of a B\"uttiker's probe to show that TRS, and hence spin polarization, is extremely sensitive to decoherence, a fact that permits folding within a single concept the qualitative effect of electron correlation, electron-phonon interaction, spin textures, and non-adiabatic effects, that have been invoked to explain the anomalous high value of the spin-orbit coupling that is apparently required to reproduce the experimental values of spin polarization\cite{Fransson2023,Kim2023,Subotnik2021}.  In fact, depending on how TRS is broken, either through a decoherence probe or by changing the boundary conditions of the helix, and hence the relative spin populations, we can use this result to reproduce important asymmetries in the enantio-specific photon absorption that have been observed in complex chiral solids and interfaces\cite{Intercalators2022,PaltierOpticalCISS2023}. The inclusion of TRS breaking is also essential in understanding the dependence of both SP and COA on the geometrical factors of the helix, e.g. its length, which has been analyzed experimentally \cite{Nakai2012,Kubo2021,Therien2022,Naskar2022,NaamanDfactor}. 
 %%%%%%%%%%%%%%%%%%%%%%%%%%%%%%%%%%%%%%%%%%%%%%%%%%%%%%%%%%%%%%%%%%%%%%%%%%%%%%%%%%%%%
 \section{Results and discussion}\label{sec2}
 %\subsection{Optical properties: the Rosenfeld's tensor}
 For optically active molecules, the chiroptical properties are usually quantified in terms of the anisotropic dissymmetry factor $g_{CD}$, which characterizes the COA. It can be written in terms of the extinction coefficients $\varepsilon_{-}(\varepsilon_{+})$ for left-(right-) polarized incident light, corresponding to the optical transition between the $n$-th and $m$-th electronic states, as: 
 \begin{equation}\label{gfactor}
g_{CD}=2\frac{(\varepsilon_{-} - \varepsilon_{+})}{(\varepsilon_{-} + \varepsilon_{+})}=\frac{4R_{nm}}{|\boldsymbol{\mu}_{nm}|^2+|\mathbf{m}_{nm}|^2}.
\end{equation}
Here, $R_{nm}$ is the rotatory strength, and $\boldsymbol{\mu}_{nm}$ and $\mathbf{m}_{nm}$ are the transition electric dipole moment, and the transition magnetic dipole moment, respectively. The rotatory strength $R_{nm}$, also known as Rosenfeld's tensor, is defined as the imaginary part of the scalar product of the two transition moments:
\begin{eqnarray}\label{GeneralRosenfeldTensor}
R_{nm} &=& \Im(\boldsymbol{\mu}_{nm}\cdot\boldsymbol{m}_{mn}).
\end{eqnarray}
This definition is the usual one in the dipole approximation, which excludes quadrupole and higher-order contributions\cite{BarronBook}. For an isotropic system, $R_{nm}$ can be written as $R_{nm}=(1/3)\left(R_{11} + R_{22} + R_{33}\right)_{nm}$, with $(R_{ii})_{nm}$ being the tensor components relative to the direction of propagation of light, e.g.,  $(R_{33})_{nm}$ refers to light incident along the $z$-direction \cite{Tinoco1964}. 

Due to the fact that organic molecules are primarily composed of low atomic number elements, atomic spin-orbit coupling (ASOC) is considered to be very weak, hence the usual definition of the rotatory strength in eq.(\ref{GeneralRosenfeldTensor}) omits the contribution of the electronic spin and any influence of ASOC on the circular dichroism  (CD) spectrum. Nonetheless, the discovery of the CISS effect presents an opportunity to investigate the role of an effective SOC that can be substantially different from the bare ASOC involved in spectroscopy. The inclusion of spin effects requires a redefinition of the   $R_{nm}$ to consider the total magnetic moment $\mathbf{J}$, adding the spin contribution $\mathbf{m}_s$ to the orbital one, i.e.,   $\mathbf{J}=\mathbf{m} + \mathbf{m}_{s}$. 
%This extension of Tinoco's model leads to a set of eigenfunctions where the SOI arises due to the confining electric field in a Rashba-type model. This characterizes the unperturbed system, which interacts with an external electromagnetic radiation with frequency  $\omega$(see Fig.\ref{Fig:HelixSystem}(a)) that induces the electronic transitions responsible for the COA. 

\begin{center}
\begin{figure*}
    \centering
\includegraphics[height=4.9cm]{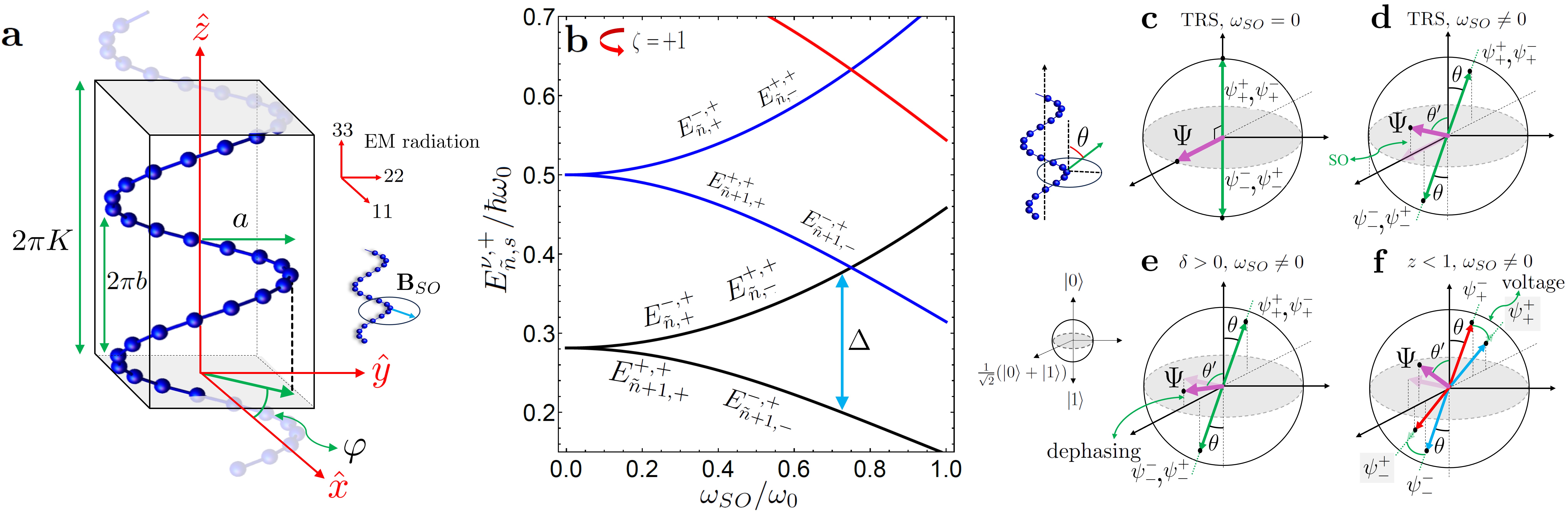}
    \caption{{\bf Helical system, energy spectrum, and the Bloch space}. {\bf a}, Structure of the helix in the molecular coordinate system $(\hat{x},\hat{y},\hat{z})$ in the presence of an external time-dependent electromagnetic field with frequency $\omega$. The structural parameters are given by the radius $a$, the pitch $2\pi b$, and the $\varphi$ angle such that $0\leq\varphi\leq 2\pi K$, with $K$ the number of turns. An effective SO magnetic field $\mathbf{B}_{SO}$ is considered perpendicular to the helix axis. {\bf b}, Energies of electrons in units of $\hbar\omega_{0}$ on an counter-clockwise helix ($\zeta=+1$) as a function of the SOC strength, with $K=2$. Each energy represents the two Kramers' pairs, i.e. $E_{\tilde{n},+}^{-,+}=E_{\tilde{n},-}^{+,+}$, and $E_{\tilde{n}+1,+}^{+,+}=E_{\tilde{n}+1,-}^{-,+}$. For convenience, we have defined frequencies $\omega_{SO}=2\alpha a/\hbar(a^2+(b/2\pi)^2)$ and $\omega_{0}=\hbar/m(a^2+(b/2\pi)^2)$\cite{Arraga2015}. {\bf c}-{\bf f} Schematic representation of the vector  $\Psi$ and the corresponding states $\psi_{s}^{\nu}$ for the enantiomer $\zeta=+1$ on Bloch sphere, showing the role of the SOC, and the dephasing (changes in $\delta$ phase) and voltage (changes in $z$ coefficient) probes. }
    \label{Fig:HelixSystem}
\end{figure*}
\end{center}
%For a helix of $K$ turns, radius $a$, and pitch $2\pi b$, the distance between consecutive turns. The electron is in the presence of the spin-orbit Rashba interaction and subjected to external electromagnetic (EM) radiation of frequency $\omega$ 
%A helix of K turns, radius $a$, and pitch $2\pi b$ can be described by the coordinates (see Fig.\ref{Fig:HelixSystem})
%\begin{equation}\label{XYZcoordinates}
%\begin{array}{ccc}
%x=a\cos{\varphi}; & y=a\sin{\varphi}; &z=b\varphi,  
%\end{array}
%\end{equation}
%in the laboratory system, with $0\ll\varphi\ll 2\pi K$. 
We then consider the  Hamiltonian for an electron with charge $e$ and rest mass $m_e$ as $\mathcal{H} = H^{0} + H'$, where $H^0$ and $H'$ refer to the non-perturbed system and the perturbation, respectively. The non-perturbed Hamiltonian includes the kinetic energy and the SOC via a Rashba-like term, $H_{SO}=\boldsymbol{\sigma}\cdot(\mathbf{p}\times\boldsymbol{\alpha})$, where $\sigma_{i}$ are Pauli's matrices, $\mathbf{p}$ is the linear momentum operator, and $\boldsymbol{\alpha}$ is a parameter of the model that includes the electric field and a coupling constant that controls the magnitude of the SOC. The SOC in a tight-binding model of a real molecule is connected to the ASOC ($\backsim$6 meV for carbon atoms) and can be enhanced by the orbital overlap between neighboring $\pi$-orbitals, resulting in an effective intrinsic SOC of the order of meV (like in carbon nanotubes \cite{Ando1999}), which depends explicitly on the chirality and geometry of the molecule \cite{Varela2016,VarelaChimia2018,TorresMedina2020,Geyer2019}. An effective SOC in a helical model can also be related to the confining molecular electrostatic field $\textbf{E}_{helix}$, which generates a SOC whose magnitude is given by $\boldsymbol{\alpha}=(e\hbar/4m_{e}^{2}c^2)\mathbf{E}_{helix}$, and where the electric field has helical symmetry\cite{Arraga2015,VarelaRashba2019}. In both scenarios, the electron is subject to an effective momentum-dependent, spin-orbit magnetic field $\mathbf{B_{SO}}$, which interacts with its spin. We assume here that $\mathbf{B_{SO}}$ is oriented perpendicular to the longitudinal axis of the helix (see Fig.\ref{Fig:HelixSystem}{\bf a}). The perturbation $H'$ includes an external electromagnetic radiation of frequency  $\omega$. 

An important ingredient of our model is that the  stationary wavefunctions $\Psi_{\tilde{n}}^{\zeta}$, necessary to describe the system to first order in the perturbation\cite{BarronBook}, can be constructed as linear combinations of  eigenstates of $H_{0}$,  $\psi_{\tilde{n},s}^{\nu,\zeta}$.  These eigenfunctions $\psi_{\tilde{n},s}^{\nu,\zeta}$ have been derived in a previous work \cite{Arraga2015} and the different quantum numbers correspond to the direction of electron propagation $\nu=+1(-1)$, the label for  spin components $s=\pm 1$, and the helicity of the state $s\nu$, i.e., the projection of the spin angular momentum in the direction of propagation. The label $\zeta=\pm 1$ corresponds to the two enantiomers, and $n$ labels  the energy channels $\tilde{n}=(n-K)/2K$, $n=1,2,3...$. Explicitly, the wavefunctions for the two enantiomers are given by:
\begin{equation}\label{LinealCombinationP}
\Psi_{\tilde{n}}^{+}=ze^{i\delta}(\psi_{\tilde{n}+1,+}^{+,+}+\psi_{\tilde{n},-}^{+,+}) -qe^{i\eta}(\psi_{\tilde{n}+1,-}^{-,+}+\psi_{\tilde{n},+}^{-,+}),
 \end{equation}
\begin{equation}\label{LinealCombinationN}
\Psi_{\tilde{n}}^{-}=ze^{i\delta}(\psi_{\tilde{n}+1,+}^{-,-}+\psi_{\tilde{n},-}^{-,-}) -qe^{i\eta}(\psi_{\tilde{n}+1,-}^{+,-}+\psi_{\tilde{n},+}^{+,-}),
 \end{equation}
where the coefficients $ze^{i\delta}$ and $qe^{i\eta}$ have been added to explore the influence of decoherence on the model by changing the phases and amplitudes of the wavefunctions. These wavefunctions satisfy the same boundary conditions used in Tinoco's landmark article \cite{Tinoco1964}, i.e., for a helix of length $2\pi K$ we have $\Psi^{\zeta}_{\tilde{n}}|_{\varphi=0}=\Psi^{\zeta}_{\tilde{n}}|_{\varphi=2\pi K}=0$. The eigenfunctions also obey Kramers' degeneracy,  which implies that states with equal helicity are degenerate. The inclusion of SOC opens a gap $\Delta$ separating states with $s\nu=+1$ from those with $s\nu=-1$, but Kramers' degeneracy is preserved unless TRS is broken (see Fig.\ref{Fig:HelixSystem}(b) for eigenvalues energy spectrum). 

\subsection{Bloch sphere and  Expectation Values}
In general, the superposition of states can be represented as Bloch vectors residing on the  Bloch sphere, where the state $\Psi_{\tilde{n}}^{\zeta}$ can be written as $\ket{\Psi}=\cos{(\theta'/2)}\ket{0} + e^{i\varphi'}\sin{(\theta'/2)}\ket{1}$. A similar representation can be made for the states $\psi_{s}^{\nu}$, where, for simplicity, we  omit the labels $\tilde{n}$ and $\zeta$.  The influence of SOC and TRS on the states on the Bloch sphere for $\tilde{n}=1$ and $\zeta=+1$ is illustrated in Figure \ref{Fig:HelixSystem}{\bf c}-{\bf f}. In the simplest case, when SOC is zero and TRS is preserved (Fig. \ref{Fig:HelixSystem}{\bf c}), the  angle of inclination $\theta$ of the spinors $\psi_{s}^{\nu}$ with respect to the vertical $z$-axis is equal to zero. Each of these states represents a pure state in the direction of $\ket{0}$, resulting in $\Psi$ being in the direction of $(\ket{0}+\ket{1})/\sqrt{2}$. The degree of polarization can be defined as $P = \sqrt{P_{x}^2 + P_{y}^2 + P_{z}^2}$, where $P_{\imath}$ represent the expectation values of $\ket{\Psi}$\cite{KesslerBook}. Hence, in this case, $P=1$, since  $P_{x}=1$, and $P_{y}=P_{z}=0$. The effect of  SOC  is depicted in Figure \ref{Fig:HelixSystem}{\bf d}. The  SO-induced magnetic field $\mathbf{B}_{SO}$ causes all the states $\psi_{\tilde{n},s}^{\nu,\zeta}$ to rotate by an angle $\theta$, and the spinor $\Psi_{\tilde{n}}^{\zeta}$ to tilt by an angle $\theta'$ with respect to the molecular axis, allowing us to interpret the impact of  SOC as a relaxation of the state vector along the direction of the SO field. This rotation results in a decrease in $P_{x}$ and an increase in $P_{z}$, causing $P$ to be less than 1.  

The inclination angle $\theta'$ of the spinor $\Psi$ can be modulated by decoherence in the presence of SOC, as shown in Figures \ref{Fig:HelixSystem}{\bf e}-{\bf f}. While a change in the $\delta$ phase (or $\eta$) does not affect the orientation of the states $\psi$, a change in the $z$ (or $q$) coefficient breaks the degeneracy of states with the same helicity (Kramers' pairs), resulting in an additional rotation of the states that adds to the $\theta$ angle, and therefore is also reflected in  $\theta'$. In other words, breaking TRS by a change of the amplitudes in the wavefunction acts as an additional contribution to the magnitude of the SO interaction, causing the angle $\theta$ of certain states to increase (or decrease) due to this effect. %{\color{red}This breakdown in the degeneracy of individual states may be related to both the asymmetry effect in the CD spectra components.QUITAR}

 \subsection{Time Reversal Symmetry, Voltage Probes and Decoherence Effects}
TRS plays an important role in this work because Onsager's reciprocity relation precludes the possibility of having non-zero SP in the linear regime unless TRS is explicitly broken. The fact that in the original experiments on electron SP in gas phase molecules, the measured polarization was very low\cite{KesslerBook} can be understood as a natural consequence of the fact that in gas phase TRS is difficult to break unless multi-photon effects are included \cite{Koch2019}. The coefficients $ze^{i\delta}$ and $qe^{i\eta}$ in Eqs.(\ref{LinealCombinationP},\ref{LinealCombinationN}) have been introduced to simulate spin-insensitive B\"uttiker's probes, allowing for the exploration of decoherence effects on the model, which introduces a disruption of symmetry in both $\nu$ directions and breaks TRS\cite{VarelaScipost2023}.  These B\"uttiker's probes can serve as dephasing probes, which preserve unitarity, where only complex phases are changed\cite{Huisman2021}, or as voltage probes, where coefficients are modified leading to non-unitary evolution\cite{DAmato1990,Pastawski2006,Ellner2014}. It can be seen that when $ze^{i\delta}=qe^{i\eta}=1$, TRS is preserved. 

In what follows, we will consider the combined influence of SOC and the breaking of TRS in our model on both the chiroptical response characterized by the anisotropic dissymmetry factor $g_{CD}$ and the spin polarization vector $\mathbf{P}=(P_x, P_y, P_z)$. We will show one of our central results that, under certain conditions, the norm of the polarization vector $P$ is a predictor of chiroptical activity.
%%%%%%%%%%%%%%%%%%%%%%%%%%%%%%%%%%%%%%%%%%%%%%%%%%%%%%%%%%%%%%%%%%%%%%%%%%%%%%%%%%%%%%%%%%%%%%%%%%%%%%%%5
 
\subsection{The Effect of SOC and TRS Breaking on the Rotatory Strength and COA}
A central element in establishing the connection between COA and the SP is to generalize the definition of the Rosenfeld tensor to include the spin contributions. The rotatory strength $R_{nm}$ (and its components) can then be expressed in the general form:
\begin{equation}
R_{nm}=\frac{3e^2}{4m_{e}c}\left[\left(\frac{a^2 b}{a^2 + b^2}\right)r_{nm} + 2\hbar S_{nm}\right],
\end{equation}
where $r_{nm}$ is a dimensionless function that contains the contribution of the orbital magnetic moment and depends on the quantum numbers $n$ and $m$ and the number of turns $K$. The spin contribution to the total orbital magnetic moment is included in the coefficient $S_{nm}$, and  SOC is hidden in the wavefunctions with which  the transition matrix elements are built 
%included in both $r_{nm}$ and $S_{nm}$ 
(see Supplementary Material for more details).   

Figure \ref{Fig:RVsSOVsDecoherence} shows the influence of the SOC strength as well as of decoherence on the rotatory strength $R$ for the transition  from the fundamental state $n=1$  to the first excited state $m=2$. Figure \ref{Fig:RVsSOVsDecoherence}{\bf a} illustrates first Tinoco's case when TRS is preserved. Breaking TRS via a dephasing probe (change in relative phase $\delta-\eta$) in the presence of SOC shifts the  spectrum, but by an amount which is independent of the SOC magnitude. On the other hand, the introduction of a voltage probe, implying a change in the amplitude $z$ in the wavefunction, results in a boundary condition determining a non-zero SP, analogous to a magnetic tip in a junction. This manifests in varying rotational strengths (see Fig.\ref{Fig:RVsSOVsDecoherence}{\bf{b}}). The tensor components $R_{11}$ and $R_{22}$ for the first transition, corresponding to incident radiation along the $\mathbf{B}_{SO}$ direction (perpendicular to the molecular axis), are the only affected by the voltage probe inducing decoherence.  This change affects the components in a way that leaves the average $R$ components almost unchanged. 

\begin{center}
\begin{figure*}[]
\centering
\includegraphics[height=5.1cm]{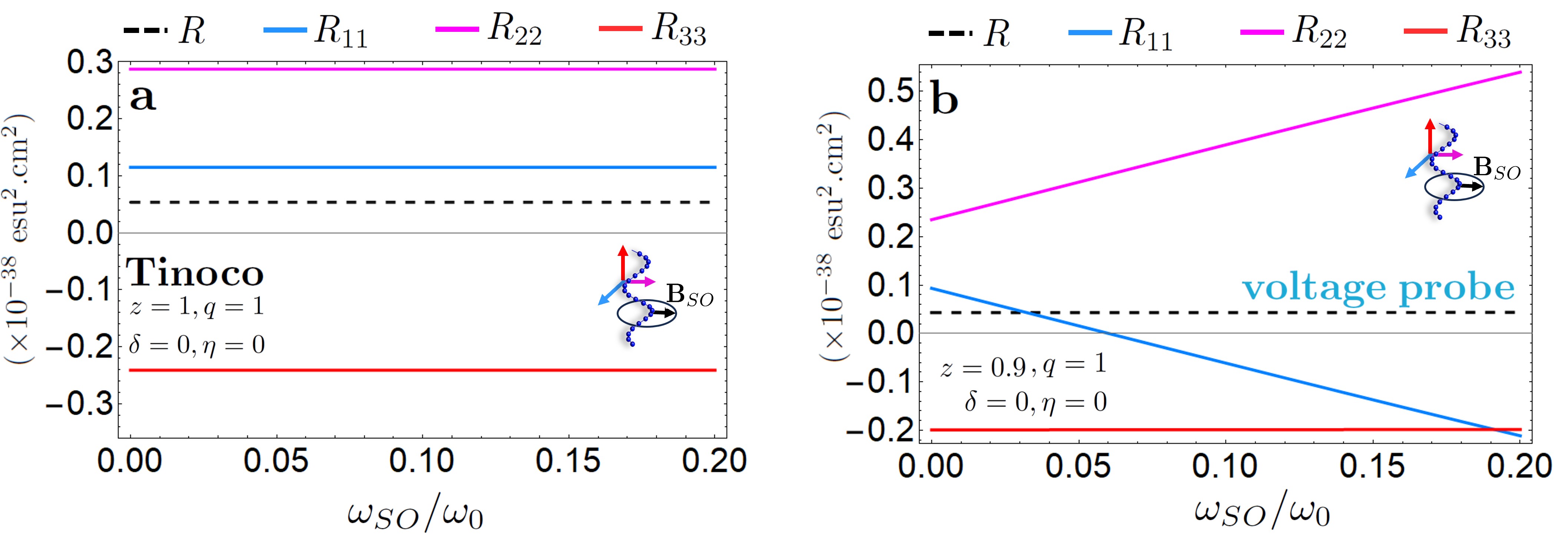}
    \caption{{\bf Rotatory strength with SO coupling and decoherence}. Rotatory strength $R$ and its components $R_{\imath\imath}$ for the transition from $n=0$ to $m=1$ as a function of the magnitude of the SO coupling.  {\bf a} shows Tinoco's case where TRS is preserved,  and {\bf b} shows the effect of decoherence induced with a voltage probe. The characteristic values of the helix used are $a=0.075$ nm, $2\pi b=3.54$ nm, and $K=1$, for positive chirality $\zeta=+1$.}
\label{Fig:RVsSOVsDecoherence}
\end{figure*}
\end{center}

%{\color{magenta}The circular dichroism (CD) spectrum of the helix for an isotropic system, for the transitions from the fundamental state $n=1$ to $m$, can be defined in terms of the rotatory strength in the form  $CD=\frac{A}{\Gamma\sqrt{\pi}}\sum_{m\neq 1}R_{1m}e^{-\left(\frac{\lambda-\lambda_{m1}}{\Gamma}\right)^2}$, with $\Gamma_{n}$ the half-width of the Gaussian, $\lambda$ the incident radiation wavelength, and $\lambda_{m1}=\lambda_{m}-\lambda_{1}$ REF.}

\begin{center}
\begin{figure*}[]
\centering
\includegraphics[height=3.7cm]{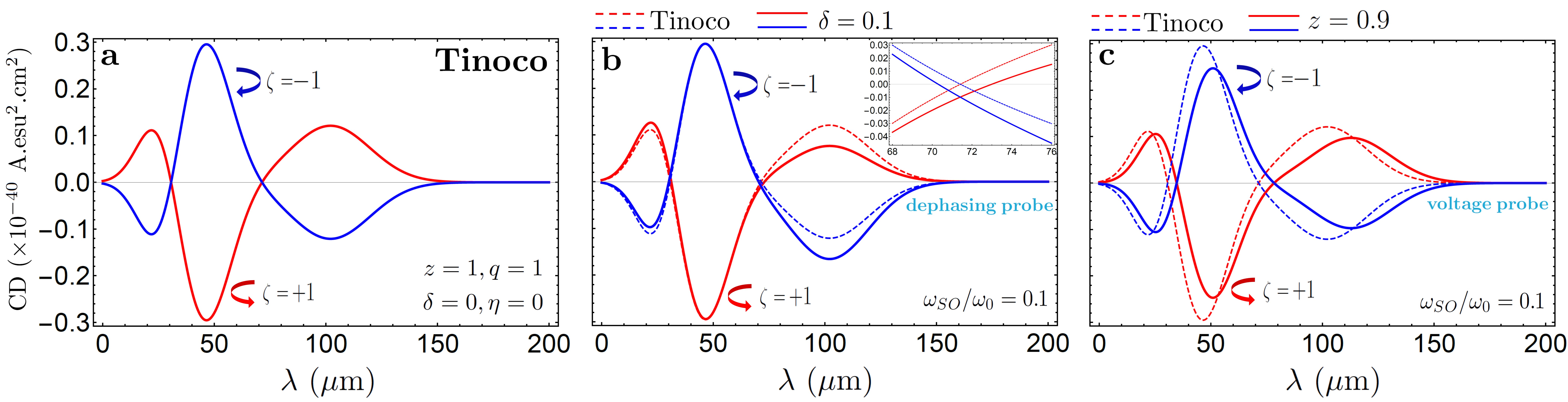}
    \caption{{\bf Circular dichroism spectrum with SOC and decoherence.} The CD spectrum for a helix is shown as a function of the incident radiation wavelength $\lambda$ for a helix with preserved TRS ({\bf a}) with SO coupling; ({\bf b}) with a dephasing probe; and ({\bf c}) with a voltage probe, for both enantiomers $\zeta$. The CD spectrum of the helix for an isotropic system, for the transitions from the fundamental state $n=1$ to $m$, was defined in terms of the rotatory strength in the form  CD$=\frac{A}{\Gamma\sqrt{\pi}}\sum_{m\neq 1}R_{1m}Exp{\left[-\left(\frac{\lambda-\lambda_{m1}}{\Gamma}\right)^2\right]}$, with $\Gamma$ the half-width of the Gaussian, $\lambda$ the incident radiation wavelength, and $\lambda_{m1}=\lambda_{m}-\lambda_{1}$\cite{NinaBook}.}
\label{Fig:CDComparation}
\end{figure*}
\end{center}

The CD spectrum of the helix for an isotropic system, for the first transitions from the fundamental state, is shown in Fig.\ref{Fig:CDComparation} when TRS is preserved. This spectrum remains the same with and without SO coupling (see \ref{Fig:CDComparation}{\bf a}), indicating that the electronic spin effect is negligible when TRS is preserved. This result can be interpreted as a manifestation of Bardarson's theorem\cite{Bardarson2008}. Both the dephasing and voltage probe effects are observable in the spectrum as a change in the magnitude of the CD intensity in graphs {\bf b} and {\bf c}, respectively. But, while dephasing is evidenced as an asymmetry in the CD spectra of the two enantiomers, the effect of the voltage probe generates the same change in the magnitude of the CD intensity for both enantiomers, in addition to a shift in the maximum absorption peaks, so that the spectra for the two enantiomers are mirror images, even in the presence of SOC.
\begin{center}
\begin{figure*}[]
\centering
\includegraphics[height=6.5cm]{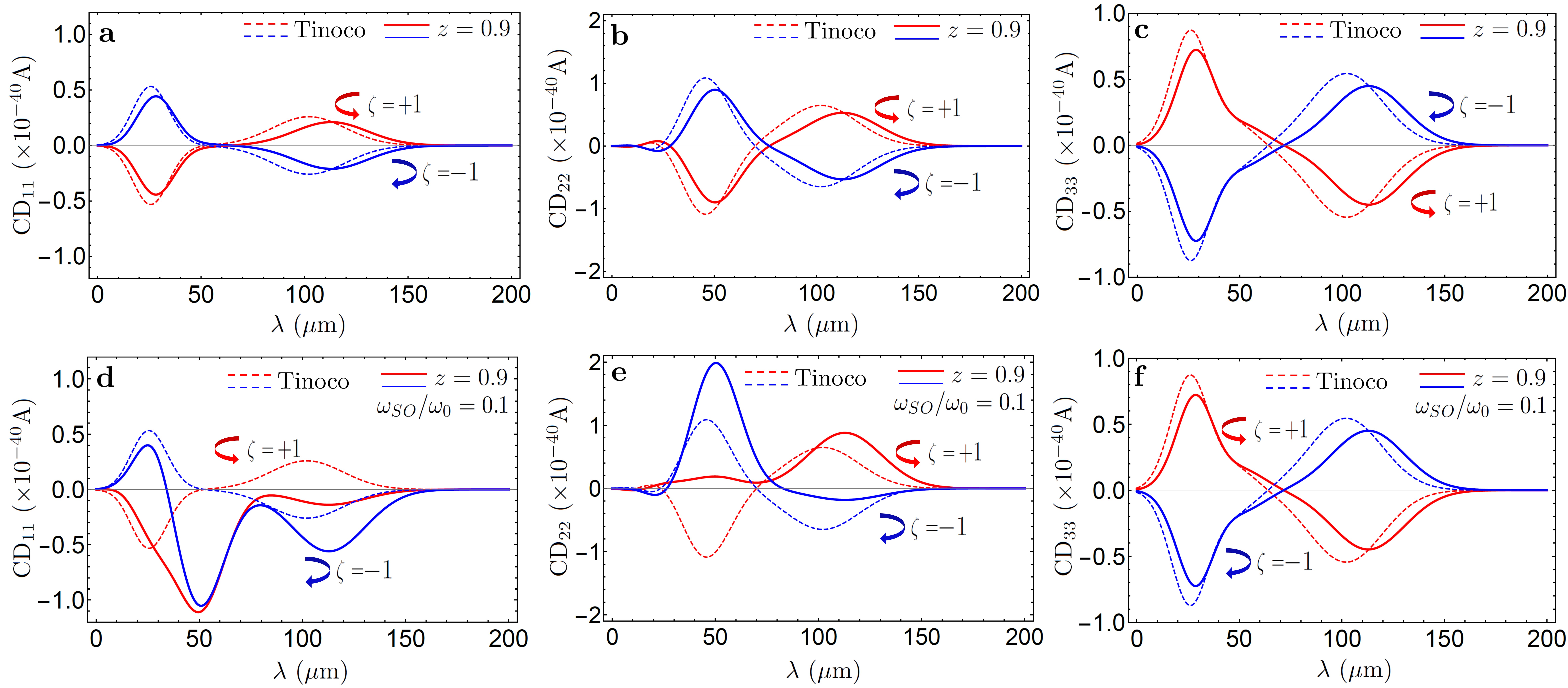}
    \caption{{\bf Circular dichroism components for oriented helix with SOC and decoherence.} Figures {\bf a}-{\bf c} illustrate the effect of the voltage probe on the CD$_{\imath\imath}$ components without SO coupling for incident radiation in the $x$, $y$, and $z$ directions, with $\imath=1,2,3$ respectively. {\bf d}-{\bf f}, Effect of decoherence coupled with SO interaction on the CD components, revealing an asymmetry in CD$_{11}$ and CD$_{22}$ components for both enantiomers (solid lines). We have used $\delta=\eta=0$, and $q=1$. }
\label{Fig:CDComponents}
\end{figure*}
\end{center}
In contrast, the impact of the voltage probe on the CD components CD$_{\imath\imath}$, relative to the direction of incidence of the radiation field,  is shown in figure \ref{Fig:CDComponents}. Inducing decoherence (without SOC) affects the spectrum in such a way that the three CD components show mirror images (\ref{Fig:CDComponents}, {\bf a}- {\bf c})
However, the coexistence of the SOC with decoherence generates an asymmetry in  CD$_{11}$ and CD$_{22}$  for the two enantiomers (see figures  \ref{Fig:CDComponents}, {\bf d}-{\bf e}), 
in contrast to  CD$_{33}$ which is symmetrical. This asymmetry is related to the orientation of the helix with respect to the direction of the incident radiation. In this way, the voltage probe simulates a magnetic probe as in the experiments, which moves the local field of the helix, changing the momentum and modifying the spin population injected into the molecule\cite{PaltierOpticalCISS2023,Intercalators2022,NaamanDfactor}.

\subsection{Spin Polarization as a Predictor of Chiroptical Activity}
As mentioned in the introductory part, one of the main motivations for this work is to establish an explicit connection between the SP induced by the CISS effect and the COA. Before entering into the details of this connection, one should emphasize a fundamental difference: while COA requires photon absorption, SP can occur as a ground-state magnetic response associated with electron transport, electron transfer, and bond polarization that occurs in the absence of external magnetic fields and also in photo-induced ET reactions. The relationship presented here is related to a very interesting interplay between SP (connected to the $P_z$ component),  and spin coherence\cite{Fay2021}. Here, we will concentrate on the simpler case of the comparison between ground state SP and COA, because it offers important insights into the connection between the two phenomena and the interpretation of recent experiments by Waldeck et al. \cite{Beratan2017} about a correlation between the length dependence of both magnitudes and the possible use of SP as a predictor of COA.

Since the COA is a scalar and the SP is a vector quantity, we have considered as a plausible scalar 
%The COA is a scalar quantity associated with the absorption intensity of polarized light by different enantiomers of a chiral molecule or material, while the SP is a vector whose components depend on the coordinate system used to describe Pauli's matrices. This consideration led us to consider as a plausible scalar 
predictor the norm  of the polarization vector $P$, which carries information about the global SP and is independent of the coordinate system. Our conjecture is also inspired by the fact that the polarization vector is simply related to the spin component of the total magnetic moment by the equation $\mathbf{m}_{s}=\it{C}\bf{P}$, where $\it{C}$ is proportional to the gyromagnetic ratio. This connection of the CISS-related response to the magnetic moment is very significant, because it emphasizes the nature of this effect as a magnetic response to electron transport. It should be stressed that in our model, transport is mimicked by changing the boundary conditions with a parameter $\epsilon$, defining the stationary states in eqs.(\ref{LinealCombinationP},\ref{LinealCombinationN}). In this case, we consider $\Psi^{\zeta}_{\tilde{n}}|_{\varphi=0+\epsilon}\neq\Psi^{\zeta}_{\tilde{n}}|_{\varphi=2\pi K}$. From this point of view, our system is equivalent to an electron in a box with adjustable boundary conditions. 

\begin{center}
\begin{figure*}[]
\centering
\includegraphics[height=5.9cm]{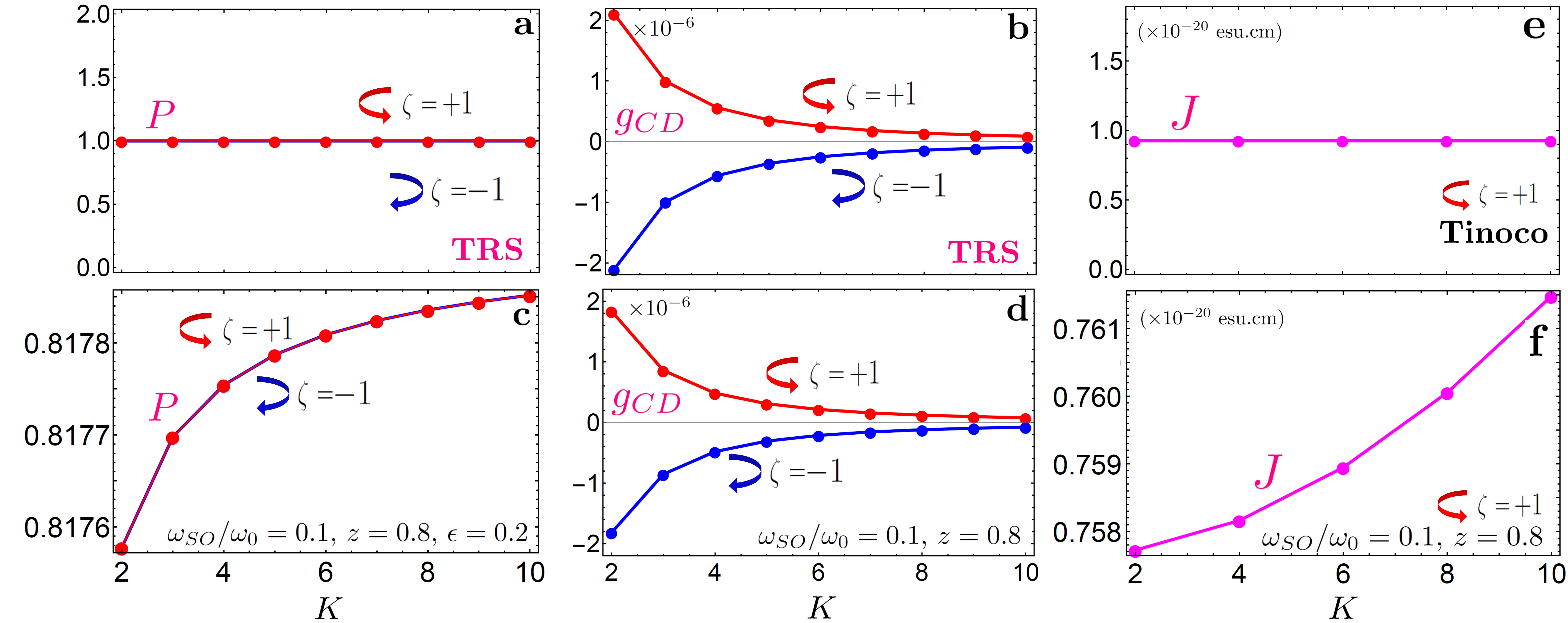}
    \caption{{\bf Length dependence of SP and COA.} Norm of the polarization vector $P$ as a function of $K$ for the fundamental state ($n=1$) ({\bf a}) when TRS is preserved, and ({\bf c}) when TRS is broken by boundary-induced decoherence, in presence of SOC. Dissymmetry factor $g_{CD}$ as a function of $K$ for the transition from the ground state with $n=1$ to the first excited state $m=2$ ({\bf b}) when TRS is preserved, and ({\bf d}) when TRS is broken using a voltage probe with SOC. Total average magnetic moment for the ground state as a function of the helix length for ({\bf e}) the Tinoco's case, and ({\bf f}) when TRS is breaking by a voltage probe in the presence of SOC. We have used $\delta=\eta=0$, and $q=1$.}
\label{Fig:Predictor} 
\end{figure*}
\end{center}

Figures \ref{Fig:Predictor} displays the graphical relationship between $P$, calculated for the ground state $n=1$,  and the dissymmetry factor $g_{CD}$, corresponding to the transition $n=1 \rightarrow m=2$ as a function of the number of turns in the helix $K$, in presence of SOC when: i) TRS is preserved (panels {\bf a} and {\bf b}); ii) TRS is broken by decoherence and changing the boundary conditions ( panels {\bf c} and {\bf d}).  Despite the fact that both phenomena have very different physical origins, there is a clear correspondence that originates from the fact that both quantities are sensitive to the CISS-induced magnetic response. That the SP component is the key predictor is also apparent from the analysis of the plot of the total average magnetic moment for the ground state, fig.\ref{Fig:Predictor}{\bf e} as a function of the length of the helix. Although this magnitude also increases with length in the presence of SOC and TRS breaking, it does not show the saturation behavior of both $P$ and $g_{CD}$, which indicates that it is dominated by the orbital angular momentum contribution. It is important to notice that the predictor, the SP, exhibits a  rather small variation as a function of length, mostly because of the constraints imposed by TRS in gas phase. The numerical proportionality between the two magnitudes is simply calculated to be $P\approx c 10^{-6}g_{CD}$. This type of relationship, one of our central results, should survive the transition to more realistic models and also to chiral interface, where the breaking of TRS arises from either non-linear conditions or because the system under consideration has open boundaries. 

\begin{center}
\begin{figure*}[]
\centering
\includegraphics[height=4.5cm]{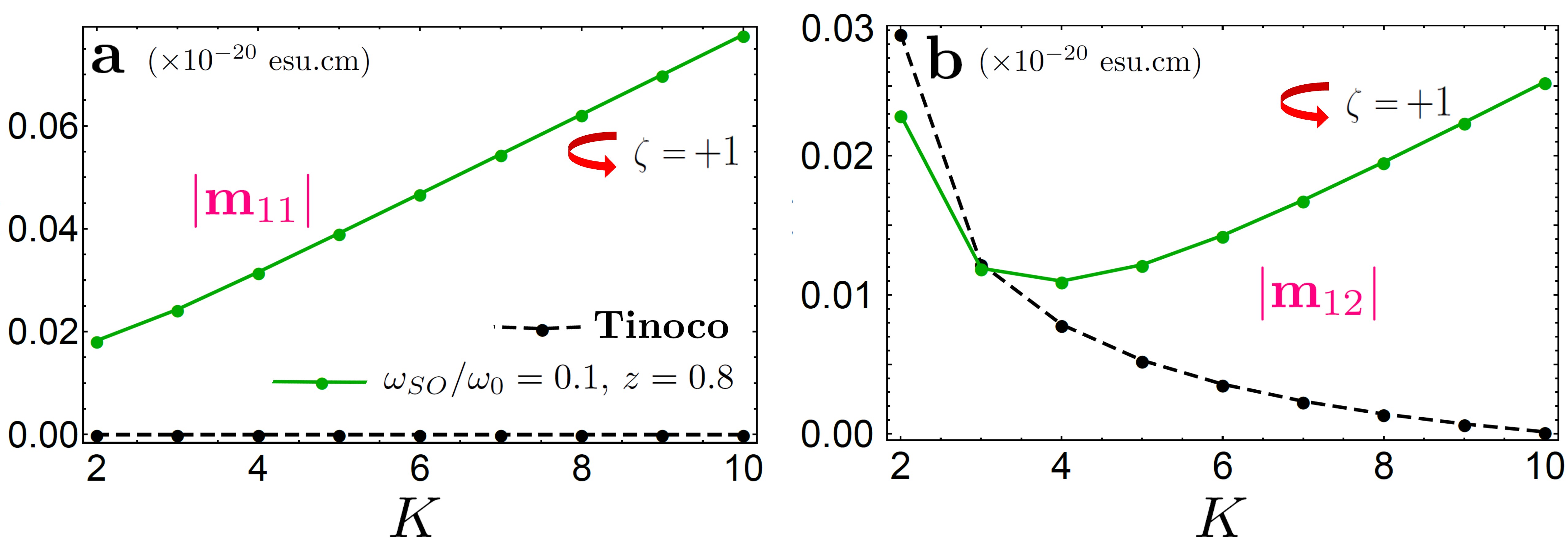}
    \caption{{\bf Length dependence of magnetic moment.} {\bf a}, Magnetic moment expectation value for the ground state as a function of the length $K$. {\bf b}, Transition magnetic dipole moment as a function of $K$ for the transition from $n=1$ to $m=2$. We have used $\delta=\eta=0$, and $q=1$.  }
\label{Fig:Mdipolar} 
\end{figure*}
\end{center}

Figures \ref{Fig:Mdipolar} $\bf a$ and $\bf b$ show the behavior of the magnetic moment expectation values for the ground state $|\mathbf{m}_{11}|$ due to the SOC, compared to spinless electrons, and the magnitude of the transition magnetic dipole moment $|\mathbf{m}_{12}|$ for the first transition, respectively. While both magnitudes have a monotonic behavior when TRS is preserved with SOC, interference effects that arise from the inclusion of decoherence can make for the non-monotone behavior of the transition moment. In any event, it is clear that these magnitudes do not have the same predicting behavior on COA as the SP, despite the fact that they depend on length.

%%%%%%%%%%%%%%%%%%%%%%%%%%%%%%%%%%%%%%%%%%%%%%%%%%%%%%%%%%%%%%%%%%%%%%%%%%%%%%%%

\section{Conclusions and Final Remarks}\label{sec13}
Using a very simple model of an electron in a helix, including the spin degree of freedom and spin-orbit coupling, we have explored a fundamental connection between optical activity and the CISS effect in chiral systems. We have also investigated the fundamental role of time-reversal symmetry in strongly modulating both the spin polarization and the chiroptical response, as well as the influence of spin-orbit coupling on the optical activity that involves a strongly asymmetric enantiomeric response, not observed in simple chiral molecules in gas phases or in solutions, but that has been found in chiral solids and interfaces. 

All our findings seem to point to a deeper connection that necessarily involves the CISS-modulated magnetic response and the fact that the chiroptical activity involves a dephasing of the electric and magnetic fields as they propagate through the chiral structure. This dephasing should be connected to the fact that the onset of the CISS effect triggers an intramolecular magnetic response that needs to be explicitly included in the spin polarization of the system. We are currently exploring this fundamental connection and also the extension of our model to describe the photon-induced spin polarization in electron transfer processes.
\backmatter

\bmhead{Supplementary information}

%%%%%%%%%%%%%%%%%%%%%%%%%%%%%%%%%%%%%%%%%%%%%%%%%%%%%%%%%%%%%%%%%%%%%%%%%%%%%%%%%%%%%%%%%%%%%%%%%%%%
\bmhead{Acknowledgments}
The authors acknowledge fruitful discussion with Prof. Jesus M. Ugalde. S.V. acknowledges the support given by the Eleonore-Trefftz-Programm and the Dresden Junior Fellowship Programme by the Chair of Materials Science and Nanotechnology at the Dresden University of Technology, and the W.M. Keck Foundation through the grant ``Chirality, spin coherence and entanglement in quantum biology."
R.G. and G.C.  acknowledge the support of the German Research Foundation (DFG)
within the project Theoretical Studies on Chirality-Induced Spin Selectivity (CU 44/55-1), and by the transCampus Research Award Disentangling the Design Principles of
Chiral-Induced Spin Selectivity (CISS) at the Molecule-Electrode Interface for Practical Spintronic Applications (Grant No. tCRA 2020-01), and Programme trans-Campus Interplay between vibrations and spin polarization in the CISS effect of helical molecules (Grant No. tC2023-03). 
E.M. acknowledges funding from project POLI17945 of USFQ and the Dresden Fellowship Programme.
V.M acknowledges the support of Ikerbasque, the Basque Foundation for Science, the German Research Foundation for a Mercator Fellowship within the project Theoretical Studies on Chirality-Induced Spin Selectivity (CU 44/55-1), and the W.M. Keck Foundation through the grant ``Chirality, spin coherence and entanglement in quantum biology."

\section*{Author contributions}
...
\section*{Competing interests}
The authors declare no competing interests.


\begin{appendices}




\end{appendices}

%\bibliography{bibliography}% common bib file

\begin{thebibliography}{99}

\bibitem{ACSNano2022} Aiello, C. D., Abendroth, J. M., Abbas, M., {\it et al}.  A Chirality-based quantum leap. {\it ACS Nano}  {\bf 16}(4), 4989-5035 (2022) DOI: 10.1021/acsnano.1c01347

\bibitem{Shi2004} Xiong, Z. H., Wu, D., Valy Vardeny, Z., \& Shi, J.  Giant magnetoresistance in organic spin-valves. {\it Nature}  {\bf 427}, 821-824 (2004) DOI: 10.1038/nature02325

\bibitem{Sun2014} Sun, D., Fang, M., Xu, X. {\it et al}. Active control of magnetoresistance of organic spin valves using ferroelectricity. {\it Nat. Commun}  {\bf 5}, 4396 (2014). DOI: 10.1038/ncomms5396

\bibitem{Dor2013} Dor, O., Yochelis, S., Mathew, S., {\it et al}.  A Chiral-based magnetic memory device without a permanent magnet. {\it Nat. Commun}  {\bf 4}, 2256 (2013). DOI: 10.1038/ncomms3256

\bibitem{Bustami2020} Al-Bustami, H., Bloom, B. P., Ziv, A., {\it et al}.  Optical multilevel spin bit device using chiral quantum dots. {\it Nano Lett.}  {\bf 20}(12), 8675-8681 (2020). DOI: 10.1021/acs.nanolett.0c03445

\bibitem{Bustami2022} Al-Bustami, H., Khaldi, O., Shoseyov, S., {\it et al}.  Atomic and molecular layer deposition of chiral thin films showing up to 99\% spin selective transport. {\it Nano Lett.}  {\bf 22}(12), 5022-2028 (2022). DOI: 10.1021/acs.nanolett.2c01953

\bibitem{Dor2017} Dor, O., Yochelis, S., Radko, A., {\it et al}.  Magnetization switching in ferromagnets by adsorbed chiral molecules without current or external magnetic field. {\it Nat. Commun}  {\bf 8}, 14567 (2017). DOI: 10.1038/ncomms14567

\bibitem{Shinto2014} Shinto, P., Prakash, M., Hagay, M., {\it et al}.  Non-magnetic organic/inorganic spin injector at room temperature. {\it Appl. Phys. Lett.}  {\bf 105}, 242408 (2014). DOI: 10.1063/1.4904941

\bibitem{Varade2018} Varade, V., Markus, T., Kiran, V., {\it et al}. Bacteriorhodopsin based non-magnetic spin filters for biomolecular spintronics. {\it Phys. Chem. Chem. Phys.}  {\bf 20}, 1091-1097 (2018). DOI: 10.1039/C7CP06771B

\bibitem{Chiesa2023} Chiesa, A., Privitera, A., Macaluso, E., {\it et al}. Chirality-induced spin selectivity: An enabling technology for quantum applications. {\it Adv. Mater.}  {\bf 35}, 2300472 (2023). DOI: 10.1002/adma.202300472

\bibitem{Santos2018} Santos J., Rivilla, I., Coss\'io, F., {\it et al}. Chirality-induced electron spin polarization and enantiospecific response in solid-state cross-polarization nuclear magnetic resonance. {\it ACS Nano}  {\bf 12}(11), 11426-11433 (2018) DOI: 10.1021/acsnano.8b06467

\bibitem{VarelaScipost2023} Varela, S., Peralta, M., Mujica, V., {\it et al}.  Spin polarization induced by decoherence in a tunneling one-dimensional Rashba model. {\it SciPost Phys. Core}  {\bf 6}, 044 (2023). DOI: 10.21468/SciPostPhysCore.6.2.044

\bibitem{Dednam2023} Dednam, W., Garc\'ia-Bl\'azquez, M. A., Zotti, L. A., {\it et al}. A group-theoretic approach to the origin of chirality-induced spin-selectivity in nonmagnetic molecular junctions. {\it ACS Nano.}  {\bf 17}(7), 6452-6465 (2023) DOI: 10.1021/acsnano.2c11410

\bibitem{Wees2019} Yang, X., van der Wal, C. H., \& van Wees, B. J. Spin-dependent electron transmission model for chiral molecules in mesoscopic devices. {\it Phys. Rev. B}  {\bf 99}, 024418 (2019) DOI: 10.1021/acsnano.2c11410

\bibitem{Wees2020} Yang, X., van der Wal, C. H., \& van Wees, B. J. Detecting chirality in two-terminal electronic nanodevices. {\it Nano Lett.}  {\bf 20}, 6148 (2020) DOI: 10.1021/acsnano.2c11410

\bibitem{Fransson2023} Fransson, J. Vibrationally Induced magnetism in supramolecular aggregates. {\it J. Phys. Chem. Lett.}  {\bf 14}(10), 2558-2564 (2023) DOI: 10.1021/acs.jpclett.3c00157

\bibitem{Subotnik2021} Wu, Y., \& Subotnik, J. E. Electronic spin separation induced by nuclear motion near conical intersections. {\it Nat Commun}  {\bf 12}, 700 (2021) DOI: 10.1038/s41467-020-20831-8

\bibitem{Volosniev2021} Volosniev, A., Alpern, H., Paltier, Y., {\it et al}.  Interplay between friction and spin-orbit coupling as a source of spin polarization. {\it Phys. Rev. B}  {\bf 104}, 024430 (2021) DOI: 10.1103/PhysRevB.104.024430

\bibitem{Peralta2020} Peralta, M., Feijoo, S., Varela, S., {\it et al}  Coherence preservation and electron-phonon interaction in electron transfer in DNA {\it J. Chem. Phys.}  {\bf 153}, 165102 (2020) DOI: 10.1063/5.0023775

\bibitem{Peralta2023} Peralta, M., Feijoo, S., Varela, S., {\it et al}  Spin-phonon coupling in a double-stranded model of DNA {\it J. Chem. Phys.}  {\bf 159}, 024711 (2023) DOI: 10.1063/5.0156347

\bibitem{Ray1999} Ray, K., Ananthavel, P., Waldeck, D. H., \& Naaman, R. Asymmetric scattering of polarized electrons by organized organic films of chiral molecules. {\it Science}  {\bf 283}, 814-816 (1999)  DOI: 10.1126/science.283.5403.814

\bibitem{Gohler2011} G\"ohler, B., Hamelbeck, V., Markus, T. Z., Kettner, M., Hanne, G. F., Vager, Z., Naaman, R., \& Zacharias, H. Spin selectivity in electron transmission through self-assembled monolayers of double-stranded DNA. {\it Science}  {\bf 331}(6019):894-7 (2011) DOI: 10.1126/science.1199339. PMID: 21330541.

\bibitem{Vager2010} Naaman, R., \& Vager, Z. Cooperative electronic and magnetic properties of self-assembled monolayers. {\it MRS Bulletin}  {\bf 35}, 429-434 (2010)  DOI: 10.1557/mrs2010.580

\bibitem{Weiss2019} Abendroth, J., Cheung, K., Stemer, D. M., {\it et al}. Spin-dependent ionization of chiral molecular films. {\it J. Am. Chem. Soc.}  {\bf 141}(9), 3863-3874 (2019) DOI: 10.1021/jacs.8b08421

\bibitem{Kettner2015} Kettner, M., G\"ohler, H., Zacharias, H., {\it et al}. Spin filtering in electron transport through chiral oligopeptides. {\it J. Phys. Chem. C}  {\bf 119}(26), 14542-14547 (2015) DOI: 10.1021/jp509974z

\bibitem{Xie2011} Xie, Z., Markus, Z., Cohen, S. R., {\it et al}. Spin specific electron conduction through DNA oligomers. {\it Nano Lett.}  {\bf 11}(11), 4652-4655 (2011) DOI: 10.1021/nl2021637

\bibitem{Nogues2011} Nogues, C., Cohen, S. R., Daube, S. S., \& Naaman, R. Electrical properties of short DNA oligomers characterized by conducting atomic force microscopy. {\it Phys. Chem. Chem. Phys.}  {\bf 6}, 4459-4466 (2004) DOI: 10.1039/B410862K

\bibitem{Kiran2016} Kiran, V., Shinto, P., Sidney, R., {\it et al}. Helicenes—A new class of organic spin filter. {\it Adv. Mater.}  {\bf 28}(10), 1957-1962 (2016) DOI: 10.1002/adma.201504725

\bibitem{Malajovich2000} Malajovich, I., Kikkawa, J. M., Awschalom, D. D., Berry, J. J., \& Samart, N. Coherent transfer of spin through a semiconductor heterointerface. {\it Phys. Rev. Lett.}  {\bf 84}(5), 1015-1018 (2015) DOI: 10.1103/PhysRevLett.84.1015

\bibitem{Min2003} Ouyang, M., \& Awschalom, D. D. Coherent spin transfer between molecularly bridged quantum dots. {\it Science}  {\bf 301}, 1074-1078 (2003) DOI: 10.1126/science.10869

\bibitem{Beratan2017} Bloom, B., Graff, B., Ghosh S., Beratan, D., \& Waldeck D.  Chirality control of electron transfer in quantum dot assemblies, {\it J. Am. Chem. Soc.}  {\bf 139}, 9038-9043 (2017) DOI: 10.1021/jacs.7b04639

\bibitem{Debabrata2013} Mishra, D., Markus, T. M., \& Naaman, R. Spin-dependent electron transmission through bacteriorhodopsin embedded in purple membrane. {\it PNAS}  {\bf 110}(37), 14872-14876 (2013)  DOI: 10.1073/pnas.1311493110

\bibitem{GhoshWaldeck2019} Ghosh, K. B., Zhang, W., Tassinari, F., {\it et al}. Controlling Chemical Selectivity in Electrocatalysis with Chiral Cuo-Coated Electrodes. {\it J. Phys. Chem. C}  {\bf 123}, 3024-3031 (2019) DOI: 10.1021/acs.jpcc.8b12027

\bibitem{Wei2006} Wei, J. J., Schafmeister, C., Bird, G., Paul, A., Naaman, R., \& Waldeck, D. H. Molecular chirality and charge transfer through self-assembled scaffold monolayers. {\it J. Phys. Chem. B}  {\bf 110}(3), 1301-1308 (2006) DOI: 10.1021/jp055145c

\bibitem{Mondal2015} Mondal, P. C., Fontanesi, C., Waldeck, D. H., \& Naaman, R. Field and Chirality effects on electrochemical charge transfer rates: spin-dependent electrochemistry.  {\it ACS Nano} {\bf 9}(3), 3377-3384 (2015) DOI: 10.1021/acsnano.5b00832

\bibitem{Zwang2018} Zwang, T. J., Tse, E. C., Zhong, D., \& Barton, J. K. A compass at weak magnetic fields using thymine dimer repair. {\it ACS Cent. Sci.}  {\bf 4}(3), 405-412 (2018) DOI: 10.1021/acscentsci.8b00008

\bibitem{Torres2020} Torres-Cavanillas, R., Escorcia-Ariza, G., Brotons-Alc\'azar, I., {\it et al}. Reinforced room-temperature spin filtering in chiral paramagnetic metallopeptides. {\it J. Am. Chem. Soc.}  {\bf 142}, 17572-17580 (2020) DOI: 10.1021/jacs.0c07531

\bibitem{GranadaCISSOptical} Ortu\~no, A., et al.  Extended enantiopure Ortho-Phenylene Ethylene (o-OPE)-based helical systems as scaffolds for supramolecular architectures: a study of chiroptical response and its connection to the CISS effect, {\it Org. Chem. Front.}  {\bf 8}, 5071-5086 (2021) DOI: 10.1039/D1QO00822F

\bibitem{PaltierOpticalCISS2023} Metzger, T., Batchu, H., Kumar, A., Fedotov, D., Oren, N., Bhowmick, D., Shioukhi, I., Yochelis, S., Schapiro, I., Naaman, R., Gidron, O., \& Paltier, Y.  Optical activity and spin polarization: the surface effect, {\it J. Am. Chem. Soc.}  {\bf 145}(7), 3972-3977 (2023) DOI: 10.1021/jacs.2c10456

\bibitem{Intercalators2022} Quian, Q., Ren, H., Zhou, J., {\it et al}. Chiral molecular intercalation superlattices. {\it Nature}  {\bf 606}, 902-908 (2022) https://doi.org/10.1038/s41586-022-04846-3

\bibitem{MujicaFieldMediated2020} Garc\'ia-Etxarri, A., Ugalde, J., S\'aenz, J., \& Mujica, V.  Field-mediated chirality information transfer in molecule-nanoparticle hybrids, {\it J. Phys. Chem. C}  {\bf 124}, 1560-1565 (2020) DOI: 10.1021/acs.jpcc.9b07670

\bibitem{FranssonTOptical2022} Das, T. K., Tassinari, F., Naaman, R., \& Fransson, J. Temperature-dependent chiral-induced spin selectivity effect: experiments and theory, {\it J. Phys. Chem. C}  {\bf 126}, 3257-3264 (2022) DOI: 10.1021/acs.jpcc.1c10550

\bibitem{Xie2022} Du, M., Liu, X., \& Xie, S. Spin-orbit coupling and the fine optical structure of chiral helical polymers. {\it J. Am. Chem. Soc.}  {\bf 142}, 17572-17580 (2020) DOI: 10.1039/D2CP01092E

\bibitem{Kubo2021} Kubo, H., Hirose, T., Nakashima, T., Kawai, T., Hasegawa, J., \& Matsuda, K. Tuning Transition electric and magnetic dipole moments: [7]Helicenes showing intense circularly polarized luminescence, {\it J. Phys. Chem. Lett.}  {\bf 12}, 1, 686-695 (2021) DOI: 10.1021/acs.jpclett.0c03174

\bibitem{Naskar2022} Naskar, S., Saghatchi, A., Mujica, V., \& Herrmann, C. Common trends of Chiral Induced Spin Selectivity and Optical Dichroism with varying helix pitch: A first principles study, {\it Isr. J. Chem.}  {\bf 62}, e202200053 (2022) DOI: 10.1002/ijch.202200053

\bibitem{Arraga2015} Medina, E., Gonz\'alez-Arraga, L., Finkelstein-Shapiro, D., Berche, B., \& Mujica, V. Continuum model for Chiral Induced Spin Selectivity in helical molecules, {\it J. Chem. Phys.}  {\bf 142}, 194308 (2015) DOI: 10.1063/1.4921310

\bibitem{Tinoco1964} Tinoco Jr., I., \& Woody, R. Optical rotation of oriented helices. IV. A free electron on a helix, {\it J. Chem. Phys.}  {\bf 40}, 160 (1964) DOI: 10.1063/1.1724854

\bibitem{Binghai2023} Li, X., Koo, J., Zhu, Z., {\it et al}. Field-linear anomalous Hall effect and Berry curvature induced by spin chirality in the kagome antiferromagnet Mn$_3$Sn. {\it Nat Commun}  {\bf 14}, 1642 (2023) DOI: 10.1038/s41467-023-37076-w

\bibitem{ButtikerOnsager2012} Jacquod, P., Whitney, R., Meair, J., \& B\"uttiker, M. Onsager Relations in Coupled Electric, Thermoelectric, and Spin Transport: The Tenfold Way, {\it Phys. Rev. B}  {\bf 86}, 155118 (2012) DOI: 10.1103/PhysRevB.86.155118

\bibitem{Bardarson2008} Bardarson, J. H. A proof of the Kramer degeneracy of transmission eigenvalues from antisymmetry on the scattering matrix, {\it J. Phys. A: Math. Theor.}  {\bf 41}, 405203 (2008) DOI: 10.1088/1751-8113/41/40/405203

\bibitem{Kim2023} Kim, K., Vetter, E., Yan, L., {\it et al}. Chiral-phonon-activated spin Seebeck effect. {\it Nat. Mater.}  {\bf 22}, 322-328 (2023) DOI: 10.1038/s41563-023-01473-9


%\bibitem{PaltierOptical2023} Metzger, T. S., Batchu, H., Kumar, A., {\it et al.}. Optical activity and spin polarization: the surface effect. {\it J. Am. Chem. Soc.}  {\bf 145}(7), 3972-3977 (2023) DOI: 10.1021/jacs.2c10456

\bibitem{Nakai2012} Nakai, Y., Mori, T., \& Inoue, Y. Theoretical and experimental studies on circular dichroism of carbo[$n$]helicenes. {\it J. Phys. Chem. A}  {\bf 116}(27), 7372-7385 (2012) DOI: 10.1021/jp304576g

\bibitem{Therien2022} Ko, C-H., Zhu, Q., Tassinari, F., \& Therien, M. Twisted molecular wires polarize spin currents at room temperature. {\it PNAS}  {\bf 119}(6), e2116180119 (2022) DOI: 10.1073/pnas.2116180119

\bibitem{NaamanDfactor} Amsallem, D., Kumar, A., Naaman, R., \& Gidron, O. Spin polarization through axially chiral linkers: Length dependence and correlation with the dissymmetry factor. {\it Chirality}  {\bf 35}, 9 (2023) DOI: 10.1002/chir.23556

\bibitem{BarronBook} Barron, L. D. {\it Molecular Light Scattering and Optical Activity} (Cambridge Univ. Press, 2004).


\bibitem{Ando1999} Ando, T. Spin-orbit interaction in carbon nanotubes, {\it J. Phys. Soc. Jpn.}  {\bf 69}, 1757-1763 (2000) DOI: 10.1143/JPSJ.69.1757

\bibitem{Varela2016} Varela, S., Mujica, V., \& Medina, E. Effective spin-orbit couplings in an analytical tight-binding model of DNA: spin filtering and chiral spin transport, {\it Phys. Rev. B}  {\bf 93}, 155436 (2016) DOI: 10.1103/PhysRevB.93.155436

\bibitem{VarelaChimia2018} Varela, S., Mujica, V., Medina, Spin-orbit coupling modulation in DNA by mechanical deformations, {\it Chimia}  {\bf 72}, 411-417 (2018) DOI: 10.2533/chimia.2018.411

\bibitem{TorresMedina2020} Torres, J., Hidalgo-Sacoto, R., Varela, S., \& Medina, E. Mechanically modulated spin-orbit couplings in oligopeptides. {\it Phys. Rev. B}  {\bf 102}, 035426 (2020) DOI: 10.1103/PhysRevB.102.035426

\bibitem{Geyer2019} Geyer, M., Gutierrez, R., Mujica, V., \& Cuniberti, G. Chirality-induced spin selectivity in a coarse-grained tight-binding model for helicene. {\it J. Phys. Chem. C}  {\bf 123}(44), 27230-27241 (2019) DOI: 10.1021/acs.jpcc.9b07764

\bibitem{VarelaRashba2019} Varela, S., Monta\~nes, B., L\'opez, F., Guillot, B., Mujica, C., \& Medina, E. Intrinsic Rashba Coupling due to Hydrogen Bonding in DNA, {\it J. Chem. Phys.}  {\bf 151}, 125102 (2019) DOI: 10.1063/1.5121025

\bibitem{KesslerBook} Kessler, J. {\it Polarized Electrons} (Springer-Verlag Berlin Heidelber, 1976 \& 1985).

\bibitem{Koch2019} Goetz, R. E., Koch, C. P., \& Greenman, L. Quantum control of photoelectron circular dichroism. {\it Phys. Rev. Lett.}  {\bf 122}, 013204 (2019) https://doi.org/10.1103/PhysRevLett.122.013204
 

\bibitem{Huisman2021} Huisman, K. H., \& Thijssen, J. M. A magnetoresistance through inelastic scattering. {\it J. Phys. Chem. C}  {\bf 125}(42), 23364-23369 (2021) DOI: 10.1021/acs.jpcc.1c06193

\bibitem{DAmato1990} D'Amato, J. L., \& Pastawski, M. Conductance of a disordered linear chain including inelastic scattering events. {\it Phys. Rev. B}  {\bf 41}, 7411 (1990) DOI: 10.1103/PhysRevB.41.7411

\bibitem{Ellner2014} Ellner, M., Bol\'ivar, N., Berche, B., \& Medina, E. Charge- and spin-polarized currents in mesoscopic rings with Rashba spin-orbit interactions coupled to an electron reservoir. {\it Phys. Rev. B}  {\bf 90}, 085305 (2014) DOI: 10.1103/PhysRevB.90.085305

\bibitem{Pastawski2006} Foa Torres, L. E. F., Pastawski H. M., \& Medina, E. Antiresonances as precursors of decoherence. {\it EPL}  {\bf 73}, 164 (2006) DOI: 10.1209/epl/i2005-10374-9

\bibitem{NinaBook} Berova, N., Nakanishi, K., \& Woody, R. W. {\it Circular Dichroism: Principles and Applications} (Wiley-VCH, 2020).

\bibitem{Fay2021} Fay, T. P. Chirality-induced spin coherence in electron transfer reactions. {\it J. Phys. Chem. Lett.}  {\bf 12}(5), 1407-1412 (2021) DOI: 10.1021/acs.jpclett.1c00009









































































\end{thebibliography}

\end{document}